\documentstyle{article}

\newcommand{\N}{{\kern+.25em\sf{N}\kern-.86em\sf{I}
\kern+.86em\kern-.25em}}

\begin{document}

\vspace*{2cm}

\begin{center}

{\LARGE The Absence of Ultralocal}

\vspace*{5mm}

{\LARGE Ginsparg-Wilson Fermions}

\vspace*{2cm}

W. Bietenholz \\

NORDITA

Blegdamsvej 17

DK-2100 Copenhagen \O

Denmark

\vspace*{1cm}

Preprint \ \ NORDITA-99/1-HE, hep-lat/9901005 

\vspace*{2mm}
 
PACS number \ 11.15.Ha

\end{center}

\vspace*{3cm}

It was shown recently by I.\ Horv\'{a}th that lattice fermions obeying the
standard form of the Ginsparg-Wilson relation cannot be ultralocal.
However, there are more general forms of the Ginsparg-Wilson relation,
which also guarantee the physical properties related to chirality,
but which are not covered by Horv\'{a}th's consideration.
Here we present a proof which applies to all
Ginsparg-Wilson fermions, demonstrating that they can only be
local in the sense of an exponential decay of their couplings,
but not ultralocal.

\newpage

A formulation of lattice fermions is characterized by some
lattice Dirac operator ${\bf D}$.
The famous Nielsen-Ninomiya No-Go theorem \cite{NN} excludes 
--- based on mild assumptions, namely Hermiticity and discrete
translation invariance --- the existence of undoubled
lattice fermions, which are chiral (in the sense that
$\{ {\bf D},\gamma_{5} \} =0$) and local (in the sense that the
couplings in ${\bf D}$ decay at least exponentially).

Recently, much attention has been attracted to an old idea
by P.\ Ginsparg and K.\ Wilson \cite{GW}, who suggested to break the chiral
symmetry on the lattice in a particularly smooth way, so that
\begin{equation} \label{GWR}
\{ {\bf D}_{x,y},\gamma_{5} \} = 2 
({\bf D} {\bf R} \gamma_{5} {\bf D})_{x,y} \ ,
\end{equation}
where the term ${\bf R}$ is non-trivial and {\em local}.
(Here we refer to a lattice of unit spacing in the $d$
dimensional Euclidean space, $x,y \in Z\!\!\! Z^ {d}$.)

In addition ${\bf R}$ commutes with $\gamma_{5}$.
This can be explained for instance by starting from the
prescription $\{ {\bf D}^{-1},\gamma_{5} \} = 2 \gamma_{5}{\bf R} =
2 {\bf R}\gamma_{5}$ (note that a local term  ${\bf R}$
doesn't shift the poles in ${\bf D}^{-1}$).
Alternatively, we can start off
by requiring invariance of the Lagrangian to $O(\varepsilon )$
for a lattice modified chiral transformation in the spirit of
Ref.\ \cite{ML}, $\bar \psi \to \bar \psi (1 + \varepsilon
[1-{\bf D}R])\gamma_{5}$, 
$\psi \to \psi (1 + \varepsilon \gamma_{5} [1-R{\bf D}])$
($R$ local; ${\bf D}R$, $R{\bf D}$ convolutions in c-space),
which leads to $\{ {\bf D},\gamma_{5} \} = {\bf D} \{ R,\gamma_{5}\}
{\bf D}$. Then we see that only the part of $R$, which commutes with
$\gamma_{5}$, contributes. This is the term that we call ${\bf R}$,
i.e. $ \{ R ,\gamma_{5} \} = 2 {\bf R} \gamma_{5}$.\\

There are three types of local lattice fermion 
formulations in the literature, 
the perfect as well as the classically perfect 
fermions \cite{GW,FPA,LP} and another formulation
by H.\ Neuberger \cite{HN}
(based on the so-called overlap formalism \cite{overlap}), 
which obey the Ginsparg-Wilson relation (GWR), eq.\ (\ref{GWR}), 
\cite{HLN,Has,Neu}. In fact, this relation preserves
the essential physical properties related to chirality
\cite{GW,HLN,Has,ML,SC}, even for chiral gauge theory \cite{MLc}.

As a virtue of the slight relaxation of the chiral symmetry
condition for ${\bf D}$, fermions obeying eq.\ (\ref{GWR})
(GW fermions) can be local in sense of an exponential decay 
of the couplings in ${\bf D}$. 
\footnote{For Neuberger fermions in QCD, locality has been
discussed in detail in Ref.\ \cite{HJL}, and it holds at least
up to moderate coupling strength. Other types of
overlap fermions have a still higher degree of locality \cite{WB,WBIH}.}
This is a great progress, but it does not mean that 
GW fermions can even be {\em ultralocal}, i.e.\ that their couplings
drop to zero beyond a finite number of lattice spacings.
The absence of ultralocal GW fermions has first
been conjectured intuitively \cite{WB}. In fact, it has been shown 
by I.\ Horv\'{a}th \cite{IH} that ultralocality
is excluded for the standard form of the GWR, which
is given by ${\bf R}_{x,y} = \frac{1}{2} \ \delta_{x,y}$.

However, the question if this is still true for
{\em any} choice of the Ginsparg-Wilson kernel
${\bf R}$ has not been answered yet, and the answer is 
not obvious at all from Horv\'{a}th's consideration.
Here we are going to prove the absence of
ultralocal solutions ${\bf D}$ for {\em all local kernels} ${\bf R}$, 
i.e.\ for {\em all Ginsparg-Wilson fermions}.\\

We start from the following observations: $(i)$ It is sufficient
to show the absence of {\em free} ultralocal GW fermions.
$(ii)$ If we can show this property in $d=2$, then ultralocal
GW fermions in all dimensions $d>2$ are ruled out as well,
because they could always be mapped on a 2d solution of the
GWR. In momentum space, such a mapping corresponds to the
restriction ${\bf D}(p_{1},p_{2}, 0, \dots ,0)$.\\

We assume Hermiticity, discrete translation invariance,
as well as invariance under reflections and exchange of the axes.
Then a general ansatz for ${\bf D}$ in $d=2$ reads
\begin{equation}
{\bf D}(p) = \rho_{\mu}(p) \gamma_{\mu} + \lambda (p) \, ,
\end{equation}
where $\lambda (p)$ is a real Dirac scalar,
whereas $\rho_{\mu}(p)$ is imaginary.
Here $\rho_{\mu}$ is odd in the $\mu$-direction and even in the other
direction, while $\lambda$ is even in both directions.
Furthermore exchange symmetry of the axes implies
$\rho_{1}(p_{1},p_{2})=\rho_{2}(p_{2},p_{1})$ and
$\lambda (p_{1},p_{2}) = \lambda (p_{2},p_{1})$.
As a consequence, the GW kernel ${\bf R}$ is a Dirac scalar,
${\bf R}$ is even, and ${\bf R} (p_{1},p_{2}) = {\bf R} (p_{2},p_{1})$.

The fermion has to be massless, and
the operator ${\bf D}$ must have the correct continuum limit, which implies
\begin{eqnarray} \nonumber
\rho_{\mu}(p) = i p_{\mu} + O(\epsilon^{3}) \, , &&
\lambda(p) \leq O(\epsilon^{2}) \, , \\
{\rm if} && p_{1}, \, p_{2} = O(\epsilon ) \, . \label{contlim}
\end{eqnarray}

We assume ${\bf D}$ --- and therefore $\rho_{\mu}$ and $\lambda$ ---
to be ultralocal, and we are going to demonstrate that such
a GW fermion does not exist.

To capture all local kernels ${\bf R} \neq 0$ we proceed in
two steps.\\

STEP 1 \\

In a first step we assume ${\bf R}$ to be {\em ultralocal}.
Then the modified operator ${\bf D}'$
\begin{eqnarray}
{\bf D}'(p) &:=& 2 {\bf R}(p) {\bf D}(p)
= \rho_{\mu}{'}(p) \gamma_{\mu} + \lambda ' (p) \nonumber \\
\rho_{\mu}{'}(p) &=& 2 r_{0} i p_{\mu} + O(\epsilon^{3}) \, , \
\lambda '(p) \leq O(\epsilon^{2}) \, , \label{Dprim}
\end{eqnarray}
is ultralocal as well (where $r_{0} := {\bf R}(p=0)$).

Now the free GWR can be written as
\begin{equation} \label{GWRfree}
- \rho_{1}{'}^{2}(p) - \rho_{2}{'}^{2}(p) + \tilde \lambda{'}^{2}(p) = 1 \, ,
\end{equation}
where $\tilde \lambda{'} (p) := 1- \lambda{'} (p)$.

A free GW fermion has to satisfy eq.\ (\ref{GWRfree}) at any
momentum $p$.
We first consider this condition only for the special case
$p_{1}=p_{2}:=q$ and look at the quantities 
\begin{eqnarray*}
\rho^{(dia)}_{n} &:=& \frac{1}{2\pi}
\int_{-\pi}^{\pi} dq \ \rho_{1}'(q,q) \exp (iqn) \quad {\rm and} \\
\ell^{(dia)}_{n} &:=& \frac{1}{2\pi}
\int_{-\pi}^{\pi} dq \ \tilde \lambda '(q,q) \exp (iqn) \quad
(n \in Z \!\!\! Z )
\end{eqnarray*}
(note that $\rho_{1}'(q,q) = \rho_{2}'(q,q)$).
They have to be ultralocal, i.e.\
confined to some finite interval $\vert n \vert \leq L_{dia}$.
We choose $L_{dia}$ so that
it is the maximal distance over which a non-trivial coupling occurs.
According to the Lemma in Ref.\ \cite{IH}, only the ``extreme'' couplings
with $n=\pm L_{dia}$ can contribute to $\rho^{(dia)}_{n}$, 
$\ell^{(dia)}_{n}$ \cite{Lem}.
From the low momentum expansion (\ref{Dprim}) we obtain
\begin{eqnarray} \nonumber
\rho^{(dia)} (q) &=& \frac{2r_{0} i \sin (L_{dia}q)}{L_{dia}} \, , \\
\ell^{(dia)} (q) &=& \cos (L_{dia}q) \, , \label{diag}
\end{eqnarray}
so that only discrete values 
\begin{equation} \label{r0dia}
r_{0} = \pm \frac{L_{dia}}{2^{3/2}}
\end{equation}
lead to a solution of the free GWR (\ref{GWRfree}) 
restricted to $p_{1}=p_{2}$.
I.\ Horv\'{a}th considered the case of a constant ${\bf R}(p) = 1/2= r_{0}$
(standard GW kernel), and he observed that there is no solution for that.
However, we see now that
the diagonal case $p_{1}=p_{2}$ is {\em not} sufficient to rule
out ultralocal GW fermions in general.
All the cases where $2^{3/2}r_{0}$ is an integer
are not covered by this consideration. \\

Of course we have exploited only a small part of condition
(\ref{GWRfree}) so far. We now take into consideration another special
case by setting $p_{2}=0$. For this ``mapping to $d=1$''
eq. (\ref{GWRfree}) simplifies to
\begin{equation} \label{map1d}
-\rho{'}_{1}^{2}(p_{1},0) + \tilde \lambda{'}^{2}(p_{1},0) = 1 \, .
\end{equation}
We repeat exactly the same procedure as in the diagonal
case, based on the Lemma in Ref.\ \cite{IH}.
In this case, we denote the maximal (and only) coupling distance of
\begin{eqnarray*}
\rho^{(1d)}_{n} &:=& \frac{1}{2\pi}
\int_{-\pi}^{\pi} dp_{1} \ \rho_{1}'(p_{1},0) \exp (ip_{1}n) 
\quad {\rm and} \\
\ell^{(1d)}_{n} &:=& \frac{1}{2\pi}
\int_{-\pi}^{\pi} dp_{1} \ \tilde \lambda '(p_{1},0) \exp (ip_{1}n)
\end{eqnarray*}
as $L_{1d}$, and eqs.\ (\ref{Dprim}) and (\ref{map1d}) 
now yield the condition
\begin{equation} \label{r01d}
2 r_{0} = \pm L_{1d} \, .
\end{equation}
We see that a number of ultralocal solutions for eq.\ (\ref{map1d})
exist. For instance, the 1d Wilson fermion solves the 1d
mapping of the standard GWR.\\

We now combine the two conditions which arise
from our two special cases of eq.\ (\ref{GWRfree}). 
Eqs.\ (\ref{r0dia}) and (\ref{r01d}) lead to the requirement
\begin{equation}
L_{dia}^{2} = 2 L_{1d}^{2}
\end{equation}
with the only solution $L_{dia}=L_{1d}=0$.
Since $\rho_{\mu}'$ is odd, $L_{dia}=0$ further implies
$\rho_{\mu}'(q,q)=0$, hence
\begin{equation} \label{R00}
{\bf R}(q,q) = 0 \, , \qquad q \in ] -\pi, \pi ] \ .
\end{equation}
Since ${\bf R}$ is ultralocal and even with respect to both axes, 
${\bf R}(p)$ can be written as
\begin{equation} \label{ulform}
\sum_{x_{1}=0}^{N} \sum_{x_{2}=0}^{N} \Gamma_{x_{1},x_{2}} 
\cos (x_{1}p_{1}) \cos (x_{2}p_{2})
\end{equation}
($N$ finite).
Exchange symmetry of the axes implies 
$\Gamma_{x_{1},x_{2}}=\Gamma_{x_{2},x_{1}}$, and from property (\ref{R00})
we infer $\, 0 = \Gamma_{x_{1},x_{2}}+ \Gamma_{x_{2},x_{1}} = 2 \Gamma_{x_{1},x_{2}}\,$
for all $x_{1},x_{2}$. However, this means ${\bf R}=0$, which contradicts
the assumptions.

This completes STEP 1, i.e. the proof of the absence of ultralocal solutions
for the class of ultralocal GW kernels ${\bf R}$.
As a side-remark we add that the
result of STEP 1 even holds if we allow for fermion doubling. \\

STEP 2 \\

In the second step we consider the case
where ${\bf R}$ decays exponentially, i.e.\
we assume it to be {\em local but not ultralocal}.
The question is, if such a kernel ${\bf R}$ exists,
that is: is it possible that
\begin{equation} \label{rat}
2 {\bf R} = \frac{\lambda}{- \rho^{2} + \lambda^{2}}
\end{equation}
is local (but not ultralocal), when $\rho_{\mu}$ and $\lambda$ are
ultralocal ?

In the ratio on the right-hand side of eq.\ (\ref{rat})
both, the numerator and the denominator are even and symmetric
under exchange of the axes. Hence both of them take the
form (\ref{ulform}), where $N$ is finite again
(because ${\bf D}$ is ultralocal).
Now we factorize all the terms with $x_{\mu} >1$ so that
only $\cos p_{1}$ and $\cos p_{2}$ occur. Furthermore
we define
\begin{equation}
c_{\mu} := 1 - \cos p_{\mu} \qquad (\mu =1,2) \, , \qquad
c_{\mu} = O(\epsilon^{2}) \, .
\end{equation}
After this factorization we can obviously express
numerator and denominator of eq.\ (\ref{rat})
as polynomials (of a finite degree) in $c_{1},\ c_{2}$.
From eq.\ (\ref{contlim}) we know
\begin{equation} \label{rrr}
- \rho^{2}(c_{1},c_{2}) = 2(c_{1} + c_{2}) + O(\epsilon^{4}) \, , \qquad
\lambda = O(\epsilon^{2}) \, .
\end{equation}

As a next step, we assume that the polynomials in ratio
(\ref{rat}) are {\em simplified maximally}. This means that the
maximal common factor of numerator and denominator is divided
off, with the condition that both preserve their form as
polynomials in $c_{1},c_{2}$ (or in other words: these
polynomials are reduced to their minimal degree).

{\em After} this simplification, we consider the denominator
and distinguish three cases:\\

{\it (a)} The simplified denominator reduces to a constant.\\

Of course, after this simplification also the numerator
is still a finite polynomial and therefore ultralocal,
hence in this case ${\bf R}$ would be ultralocal.
This means that we are actually not in the class of
GW kernels that we want to consider in STEP 2; this
case has already been discussed (and ruled out) in STEP 1.\\

{\it (b)} The simplified denominator vanishes at $c_{1}=c_{2}=0$.\\

Then the same must be true for the numerator,
since ${\bf R}(p)$ must be regular.
For small momenta, we call the order of the denominator
$O(\epsilon ^{2k})$, and that of the numerator
$O(\epsilon ^{2\bar k })$ (where $k,\ \bar k$ are natural
numbers and $\bar k \geq k \geq 1$) so that 
${\bf R}(c_{1},c_{2}) = O(\epsilon ^{2 (\bar k -k)})$.
Now we take $(\bar k -k +1)$
derivatives of ${\bf R}$ with respect to $c_{1}$ or $c_{2}$.
The result will diverge at $p=(0,0)$ (the situation, where
such a derivative vanishes does not belong to case {\it (b)}).
Therefore ${\bf R}$ is {\em not} analytical in momentum space,
and hence it is non-local.
So in this case we are not dealing with a GW fermion.\\

{\it (c)} The simplified denominator is momentum dependent, and it
does not vanish at $c_{1}=c_{2}=0$.

Without loss of generality we can assume the denominator
to take the form $1 + O(\epsilon^{2})$.\\

At first glance, this case seems to allow for many
ultralocal solutions. As a simple example, one could
set $\lambda = -\rho^{2}$. However, such a term $\lambda$
does not avoid doubling, hence in this case --- and
only in this case --- the condition that our
GW fermion is {\em free of doubling} is crucial.
Technically it means that ${\bf D}$ has no zeros in the
Brillouin zone $ ]-\pi , \pi ]^{2}$, except for the physical
one at $p_{1}=p_{2}=0$.

In case {\it (c)}, in the above simplification a momentum
dependent factor was divided off. We call this factor $K$,
and $-\rho^{2} = K \, S$, $\lambda = K \, T$, where $K,S,T$
are all polynomials in $c_{1},c_{2}$.

We now consider the possible forms of $K$ and $S$.
Let us first go back to the terms $\rho_{\mu}(p_{1},p_{2})$.
They can be factorized in the same way as we treated eq.\ (\ref{rat}), 
and we arrive at the form
\begin{eqnarray} \nonumber
\rho_{1}(p_{1},p_{2}) &=& i \sin p_{1} \, F(c_{1},c_{2}) \, , \\
\rho_{2}(p_{1},p_{2}) &=& i \sin p_{2} \, F(c_{2},c_{1}) \, , \label{r1}
\end{eqnarray}
where $F(c_{1},c_{2}) = 1 + O(\epsilon^{2})$ is once
more a polynomial. This implies
\begin{equation}  \label{r2}
- \rho^{2} = K \, S = c_{1}(2-c_{1}) F(c_{1},c_{2})^{2} 
+ c_{2}(2-c_{2}) F(c_{2},c_{1})^{2} \, .
\end{equation}

Since $S = 1+O(\epsilon^{2})$,
we see from eq.\ (\ref{rrr}) that $K$ can be written as
\begin{equation} \label{KX}
K = c_{1} X(c_{1},c_{2}) + c_{2} X(c_{2},c_{1}) \, ,
\end{equation}
where $X(c_{1},c_{2}) = 2 + O(\epsilon^{2})$.
$X$ is again a polynomial in $c_{1},c_{2}$,
and it is strictly forbidden that it contains
any factor $(2-c_{1})$ or $(2-c_{2})$. (Otherwise
this would also be a factor of $\lambda$, and
then doubling occurs at momentum $(p_{1},p_{2})
= (\pi ,0 )$ resp. $(0 ,\pi )$.)

On the other hand, such factors are allowed in $S$.
We finally decompose $S$ as
\begin{equation} \label{SY}
S(c_{1},c_{2}) = (2-c_{1})^{n_{1}} (2-c_{2})^{n_{2}} 
Y (c_{1},c_{2}) \, .
\end{equation}
where $n_{1},n_{2} \in \N_{0}$. 
This decomposition is done such that $Y$ does not
contain any factors $(2-c_{1})$ or $(2-c_{2})$;
all these factors are extracted, hence $n_{1},n_{2}$ are maximal.
Of course, due to exchange symmetry in the axes we know that
\begin{equation} \label{esym}
n_{1} = n_{2} \, .
\end{equation}

Combining eqs.\ (\ref{KX}) and (\ref{SY}) we arrive at
\begin{eqnarray}
K \, S &=& c_{1} (2-c_{1})^{n_{1}} (2-c_{2})^{n_{2}}
X(c_{1},c_{2}) Y(c_{1},c_{2}) \nonumber \\
&+& c_{2} (2-c_{1})^{n_{1}} (2-c_{2})^{n_{2}}
X(c_{2},c_{1}) Y(c_{1},c_{2}) \, .
\end{eqnarray}
We recall that $X$ and $Y$ do not contain any factors
$c_{\mu}$ or $(2-c_{\mu})$, due to the above decompositions.
Together with eqs.\ (\ref{r1}), (\ref{r2}) we obtain
\begin{equation}
F(c_{1},c_{2})^{2} = (2-c_{1})^{n_{1}-1} (2-c_{2})^{n_{2}}
X(c_{1},c_{2}) Y(c_{1},c_{2}) \, .
\end{equation}
Since $F(c_{1},c_{2})$ is a polynomial itself,
we conclude that $n_{1}$ must be odd, whereas
$n_{2}$ must be even.

This contradicts eq.\ (\ref{esym}), and therefore
case {\it (c)} is excluded as well. $\ \Box$

\vspace{1cm}

Now we have completed the general proof that GW fermions cannot be 
ultralocal.
\footnote{W. Kerler suggests that the GWR should generally take the form
$\, \{ {\bf D}, \gamma_{5} \} = 2 {\bf R D} \gamma_{5} {\bf D} \,$
instead of eq.\ (\ref{GWR}) \cite{WK}. Of course, our proof applies to that 
formulation too.}

In view of practical applications,
this result means that we cannot simulate fermions obeying the
GWR as formulated in the infinite volume.
In finite volume with certain boundary conditions, 
the GWR --- with these boundary conditions implemented --- may hold,
but this requires the coupling over all distances in the given volume, 
which is inconvenient.
What one can work on is a very fast exponential
decay of the couplings \cite{FPA,WB,WBIH}. In order
to construct an overlap fermion with a high level of locality,
it turned out to be useful to start from a short-ranged approximate
GW fermion, which is then inserted into the overlap formula.

As a criterion for the quality of a short-ranged approximate
free GW fermion, we can insert it into the GWR and solve for
${\bf R}$. This term is only a pseudo-GW kernel, since it has got to
be non-local, according to our result. Indeed, if we insert for
instance the Wilson fermion, the resulting term decays as
${\bf R}_{x,y} \sim 1/ (4 \, \vert x -y \vert^{4})$ in $d=2$, and
${\bf R}_{x,y} \sim 1/ (1.6 \, \vert x -y \vert^{6})$ in $d=4$.
For comparison, a truncated perfect free hypercube fermion
(with couplings inside a unit hypercube on the lattice)
provides a better approximation to a GW fermion, and the corresponding
pseudo-GW kernel decays much faster \cite{Dubna}:
${\bf R}_{x,y} \sim 1/ (290 \, \vert x -y \vert^{4})$ in $d=2$, resp.
${\bf R}_{x,y} \sim 1/ (120 \, \vert x -y \vert^{6})$ in $d=4$.\\

To summarize, we repeat that we are dealing with a new variant
of a No-Go theorem for lattice fermions. The well-known
Nielsen-Ninomiya theorem excludes locality if the fermion obeys 
$\{ {\bf D},\gamma_{5} \} =0$. {\em If we relax this condition
to the GWR, then locality is possible, but ultralocality
still not.} We have demonstrated this for a all
GW kernels ${\bf R}$, in any dimension $d \geq 2$,
and therefore for all GW fermions.\\

{\small\sf
I would like to thank G.\ Ananos and I.\ Horvath
for useful comments. The first version of this note
only presented STEP 1. Meanwhile, the result of
that first part has been confirmed in Ref.\ \cite{IH2}.
Finally I thank the organizers of the winter workshop in 
Benasque, where part of this work was done.}


\begin{thebibliography}{20}
\bibitem{Lem} Consider two functions $A_{n}$, $B_{n}$ 
$(n \in Z \!\!\! Z )$, which obey in momentum space
$A(k)^{2} + c B(k)^{2}=1$ (for some $c\neq 0$).
If $L$ is the maximal range of $A_{n}$, $B_{n}$
(i.e.\ $L$ is the minimal integer so that $A_{n}=B_{n}=0$
for all $\vert n \vert > L$), then the Lemma states that
$A_{n}=B_{n}=0$ also if $\vert n \vert < L$.
This can be shown by induction in coodinate space \cite{IH}.

This Lemma applies to 
$ -2 \rho^{(dia)}(q)^{2} + \ell^{(dia)}(q)^{2}=1$,
as well as eq. (\ref{map1d}).
\end{thebibliography}
\end{document}